\documentclass[%
 reprint,
superscriptaddress,
 amsmath,amssymb,
 aps,
prb,
]{revtex4-1}

\usepackage{xcolor}
\usepackage{graphicx}% Include figure files
\usepackage{dcolumn}% Align table columns on decimal point
\usepackage{bm}% bold math
\newcommand{\cao}{\c c\~ao\ }

\newcommand{\ii}{\'\i}
\begin{document}

%\preprint{APS/123-QED}

\title{Dynamical regimes of ultrafast kinematic vortices in the resistive state of mesoscopic superconductors}% Force line breaks with \\ 

\author{A.\ Presotto}
 \email{alice.presotto@unesp.br} %\altaffiliation {}%Lines break automatically or can be forced with 
 \affiliation{%\textcolor{red}{easily}
 UNESP, Faculdade de Engenharia de Ilha Solteira, Departamento de F\'{\i}sica e Qu\'{\i}mica
 \
}%
 
 \author{E.\ Sardella}%
\email{edson.sardella@unesp.br}
 \author{A.\ L.\ Malvezzi}%
\email{andre.malvezzi@unesp.br}
\affiliation{%
 UNESP, Faculdade de Ci\^encias, Departamento de F\'{\i}sica
 \
}%

\author{R.\ Zadorosny}%
\email{rafael.zadorosny@unesp.br}
\affiliation{%
 UNESP, Faculdade de Engenharia de Ilha Solteira, Departamento de F\'{\i}sica e Qu\'{\i}mica
 \
}%

\date{\today}% It is always \today, today,

\begin{abstract}
A superconducting state coexists with static electrical fields under the formation of phase-slips and kinematic vortices ($kV$'s). Besides that, such a resistive state installed in the superconductor is not desirable in many applications. Then, it is essential to know the ultrafast $kV$ dynamics for controlling the fluxonics of the system. Thus, in this work, we studied the dynamics of the $kV$'s in a mesoscopic superconductor under the influence of external magnetic fields ($H$) and transport currents ($J_{tr}$). As a result, the $kV$ dynamics are affected by increasing $H$. With those significant changes, it was possible to build a $J_{tr} (H)$ phase diagram. There, in between the Meissner and the normal states, the $kV$'s present three distinct behaviors. For high fields, the vortices behave as Abrikosov-like ones, with velocities two orders of magnitude lower than those at low fields regime. Besides that, one demonstrates how the interplay between $J_{tr}$ and the shielding currents, its controlled by $H$, allowing for some quantitative predictions of boundaries in the phase diagram. Additionally, for $H$'s where only a $kV$ is nucleated, for tenths of picoseconds, a surface barrier effect acts over the instantaneous velocity of the $kV$. 
%\begin{description}
%\end{description}
\end{abstract}

\pacs{Valid PACS appear here}% PACS, the Physics and Astronomy
\keywords{TDGL, vortex, mesoscopic, resistive state}%Use showkeys class option if keyword
\maketitle

%\tableofcontents

\section{\label{sec:level1}Introduction}
%\textcolor{red}{test}

Transport current flowing through nanoscopic 
superconductors can induce resistive states caused by the appearance of phase-slips (PS) \citep{Andronov,Berdiyorov-2009,Berdiyorov-Pri,Belkin-2015,Chen-2014,Petkovic-2016} or by the motion of kinematic vortices ($kV$).\citep{Andronov,Berdiyorov-Pri,Berdiyorov-2009-80,Barba-Ortega-2017} The former originates from the breaking of the Cooper pairs when the depairing current is locally reached, and the latter one is formed by local perturbation of the superconducting currents.\citep{Andronov} Such phenomenon occurs even in the absence of external magnetic fields and disregarding the self-field generated by the currents.\citep{Andronov} In both situations, i.e., in the PS and $kV$, the local superconducting order parameter, $\psi$, is momentarily quite vanished, oscillating between its maximum and minimum values.\citep{Berdiyorov-Pri,Ivlev-1984} As a consequence, an oscillating voltage is induced across the sample with a frequency of the order of terahertz.\citep{Berdiyorov-Pri} In this way, the superconducting state can coexist with a static electric field.\citep{Berdiyorov-Pri,Berdiyorov-2009-80,Ivlev-1984} 

The $kV$'s have interesting features in comparison with Abrikosov-like vortices ($AbV$). Their velocities are in between one and two orders of magnitude larger than that one presented by $AbV$.\citep{Berdiyorov-Pri} Consequently; the $kV$'s are anisotropic with an elongated core shape.\citep{Andronov} Besides that, the $kV$'s move through the material under the Lorentz Force, and their velocities are directly related to the distribution of currents through the sample, which means that the more uniform the arrangement of the currents is, the faster the vortex will move.\citep{Berdiyorov-Pri}

However, external magnetic fields induce an asymmetric distribution of the superconducting current density, $\textbf{J}_{s}$, influencing the dynamics of the $kV$'s and decreasing their velocities.\citep{Berdiyorov-Pri} Additionally, resistive states induced by transport currents cannot be desirable since they reduce the sensitivity of devices such as the superconducting single-photon detectors.\citep{Berdiyorov-2012,Rosticher-2010,Zhang-2016,Dorenbos-2008} On the other hand, there is a dual relationship between PS and Josephson junctions which can allow the design of standard currents devices as well.\citep{Mooij-2005,Giblin-2012,Kaneko-2016,FENTON-2016}

In this work, we study the dynamics of $kV$'s in a sample with a central constriction under the influence of externally applied magnetic fields with different amplitudes. As will be discussed in detail below, the constriction produces a non-uniform distribution of the transport current across the sample, allowing for the formation of kinematic vortex-antivortex ($V-aV$) pairs. In this sense, its effect is
similar to narrowing the current leads. Differently, it precludes the appearance of two types of $V-aV$ dynamics\citep{Berdiyorov-Pri}, favoring instead only one of these regimes. 
A current-field diagram was built showing the phases of the $kV$ dynamics. Besides, we also show that surface-barrier effects can influence the motion of a $kV$. 

This work is organized as follows. In Section \ref{sec:level2}, the theoretical framework is briefly introduced. Our approach relies on the generalized Ginzburg-Landau equation (GTDGL for short). Results and discussions appear in Section \ref{sec:level3}. First we present the current-voltage ($I-V$) characteristic. Second, we analyze the frequency and amplitude of the time-voltage characteristic and the we determine a $J_{tr}-H$ phase diagram delimiting the Meissener, $V-aV$, only V, and $AbV$ states; a theoretical foundation for this diagram is also provided. Here, $j_{tr}$ stands for the $dc$ applied current density and $H$ for the external applied magnetic field. Thirdly and last, we exhibit surface barrier effects on the motion of $kV$'s. Finally, the concluding remarks are presented in Section \ref{sec:level4}.

\section{\label{sec:level2}Theoretical Model}

\begin{figure}[h!]
    \centering
    \includegraphics[width=7.0cm]{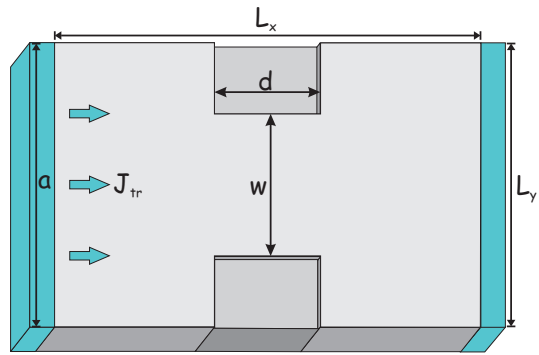}
    \caption{Representation of the simulated sample, where the transport current is applied through the metallic contacts. See main text for details.}
    \label{fig:amostra}
\end{figure}

The objective of this work is to analyze a superconducting system under
transport currents and applied magnetic fields. To that purpose, two surface defects of a superconductor with lower $T_c$ are considered to form a constriction in the central portion of a superconducting tape. The $dc$ applied current density, $J_{tr}$, is injected through the metallic contacts, as can be seen in Fig.~\ref{fig:amostra}. Then, we simulate the superconducting sample by solving the GTDGL equation\citep{Kramer-1978,RWatts-1981} which is given by

\begin{equation}
\begin{aligned}
 {\frac{u}{\sqrt{1+\gamma^2|\psi|^2}}\left({\frac{\partial}{\partial{t}}+i\varphi+\frac{\gamma^2}{2}\frac{\partial{|\psi|^2}}{\partial{t}}}\right)\psi} \\
=-(-i\mbox{\boldmath $\nabla$}-\mathbf{A})^2\psi+\psi(g(\textbf{r})-|\psi|^2),  
\end{aligned}
\label{eq:1}
\end{equation}
where $\gamma=2\tau_{E}\psi_{0}/\hbar$, $\tau_{E}$ being the characteristic time of the inelastic collision of the normal electrons, $g(\textbf{r})$ is an \textit{ad-hoc} function related to the local value of $T_c$, i.e., in the defects $g(\textbf{r})=0$ (superconductor with lower $T_c$) and in the superconducting matrix $g(\textbf{r})=1$. The parameter $u = 5.79$ is determined from quantum-mechanical considerations. \citep{Kramer-1978}

Eq.~(\ref{eq:1}) is coupled with the scalar potential equation,

\begin{equation}
    \nabla^2\varphi=\mbox{\boldmath $\nabla$} \cdot\mathbf{J}_{s},
\label{eq:2}
\end{equation}
where $\textbf{J}_{s}$ is the density of superconducting current. This equation can be derived from the continuity equation by assuming that there is no charge accumulation, that is, $\mbox{\boldmath $\nabla$}\cdot\mathbf{J}=0$, where 
$\mathbf{J}=\mathbf{J}_s+\mathbf{J}_n$ is the total and $\mathbf{J}_n=-\mbox{\boldmath $\nabla$}\varphi$ is the normal current density, respectively.

Eqs.~(\ref{eq:1}) and (\ref{eq:2}) are already written in a normalized form, where the length is expressed in units of the coherence length $\xi$, the temperature $T$ in units of $T_c$, the time in units of the GL characteristic time  $\tau_{GL}=\pi\hbar/k_BT_cu$, the magnetic field in units of the upper critical field $H_{c2}$, the electrostatic potential in units of $\varphi_0=\hbar/2e\tau_{GL}$, the vector potential in units of $H_{c2}\xi$, current density in units of $J_0=c\sigma\hbar/2et_{GL}$ ($\sigma$ is the electrical conductivity in the normal state), and the order parameter in units of $\psi_0 = \sqrt{|\alpha|/\beta)}$ (the order parameter in the Meissner state), where $\alpha$ and $\beta$ are the GL phenomenological constants. We have solved eqs.~(\ref{eq:1}) and (\ref{eq:2}) by using the link-variable method.\cite{Gropp-1996} 

The geometry considered in the present work is illustrated in Fig.~\ref{fig:amostra}. It is constituted of a mesoscopic superconducting stripe with a constriction in the middle, and two attached normal metallic contacts. The Neumann boundary condition ${\bf n}\cdot\mbox{\boldmath $\nabla$}\varphi=0$ is taken at all sides, except at the metallic contacts where we use ${\bf n}\cdot\mbox{\boldmath $\nabla$}\varphi=-J_{tr}$. For the order parameter we use the boundary condition ${\bf n}\cdot(i\mbox{\boldmath $\nabla$}+{\bf A})\psi = 0$ on all sides, except at the metallic contacts where we employ the Dirichlet boundary condition $\psi=0$. Here ${\bf n}$ is a unit vector normal to all sides of the sample pointing outward.

\section{\label{sec:level3} Results and Discussion} 

\subsection{\label{sec:level3a} Current-Voltage Characteristic Curves and Dynamics}

The simulations were carried out for a sample with lateral sizes of $L_x=12\xi$ and $L_y=8\xi$. Additionally, the constriction sizes are fixed at  $w=4\xi$ and $d=3\xi$ and the width of the electrical contacts is considered as $a=L_y$. Although we consider $T = 0$, the GDTGL equations can be applied to superconductors at $T\geq 0.5 T_c$,\citep{Petkovic-2016} 
or equivalently, the sizes can  be properly adjusted according to $\xi(T)=\xi(0)/\sqrt{1-T/T_c}$.\citep{Berdiyorov-Pri} For instance, for $T=0.96T_c$ and $\xi(0)=10$ nm , we have $\xi=50$ nm for thin Nb films. This gives $L_x=600$ nm and $L_y=400$ nm, values which were used in Ref.~\onlinecite{Berdiyorov-Pri}. In addition, while $J_{tr}$ is varied the external field $\mathbf{H}$, applied perpendicular to the surface of the sample, is maintained fixed. Besides that, the behavior of the systems are analyzed by the $I-V$ characteristic curves, distributions of the $\textbf{J}_{s}$, intensity of $\psi$ among other physical quantities.

\begin{figure}[h!]
    \centering
    \includegraphics[scale=0.37]{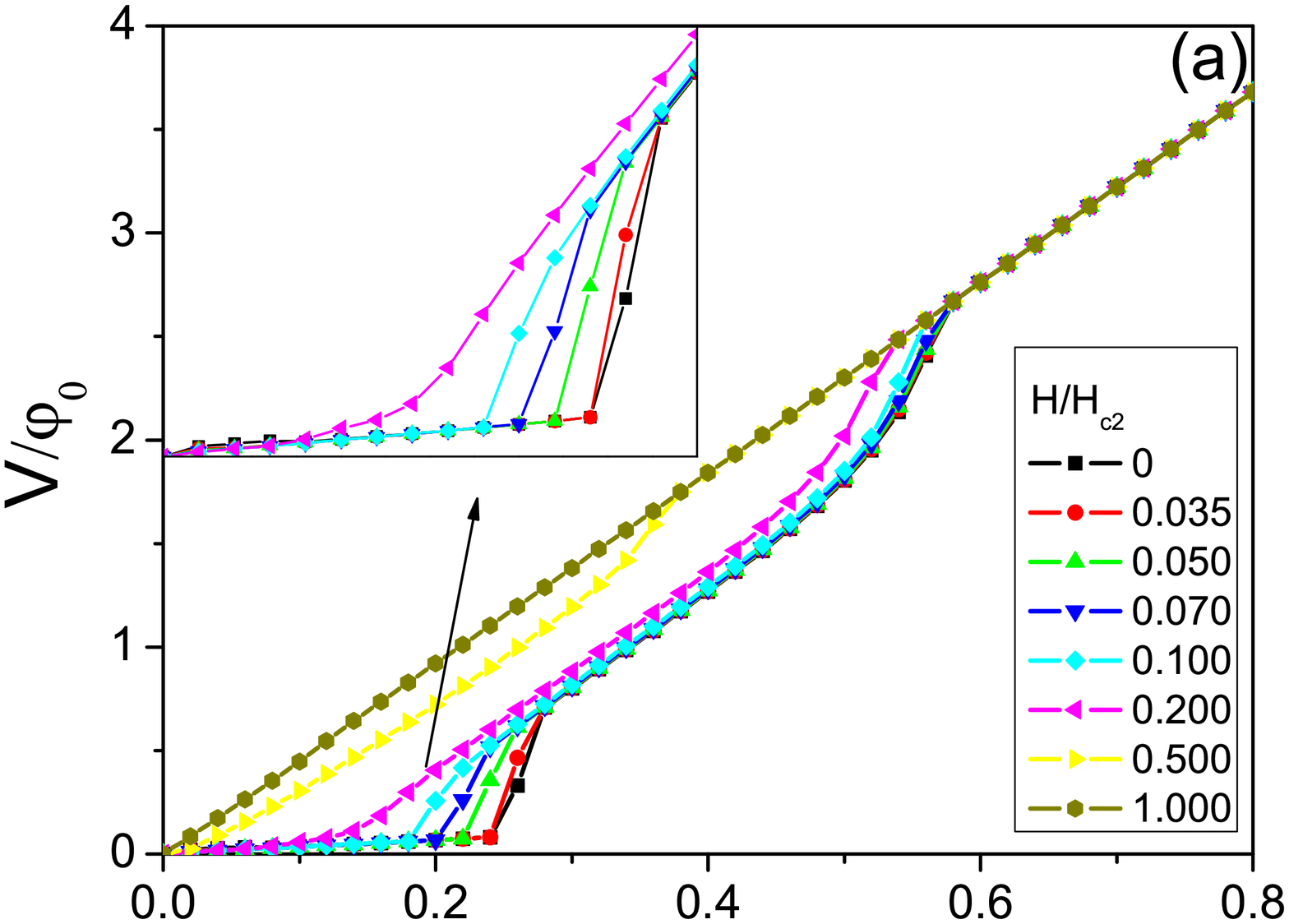}\\
    \vspace{0cm}\hspace{-0.06cm}\includegraphics[scale=0.37]{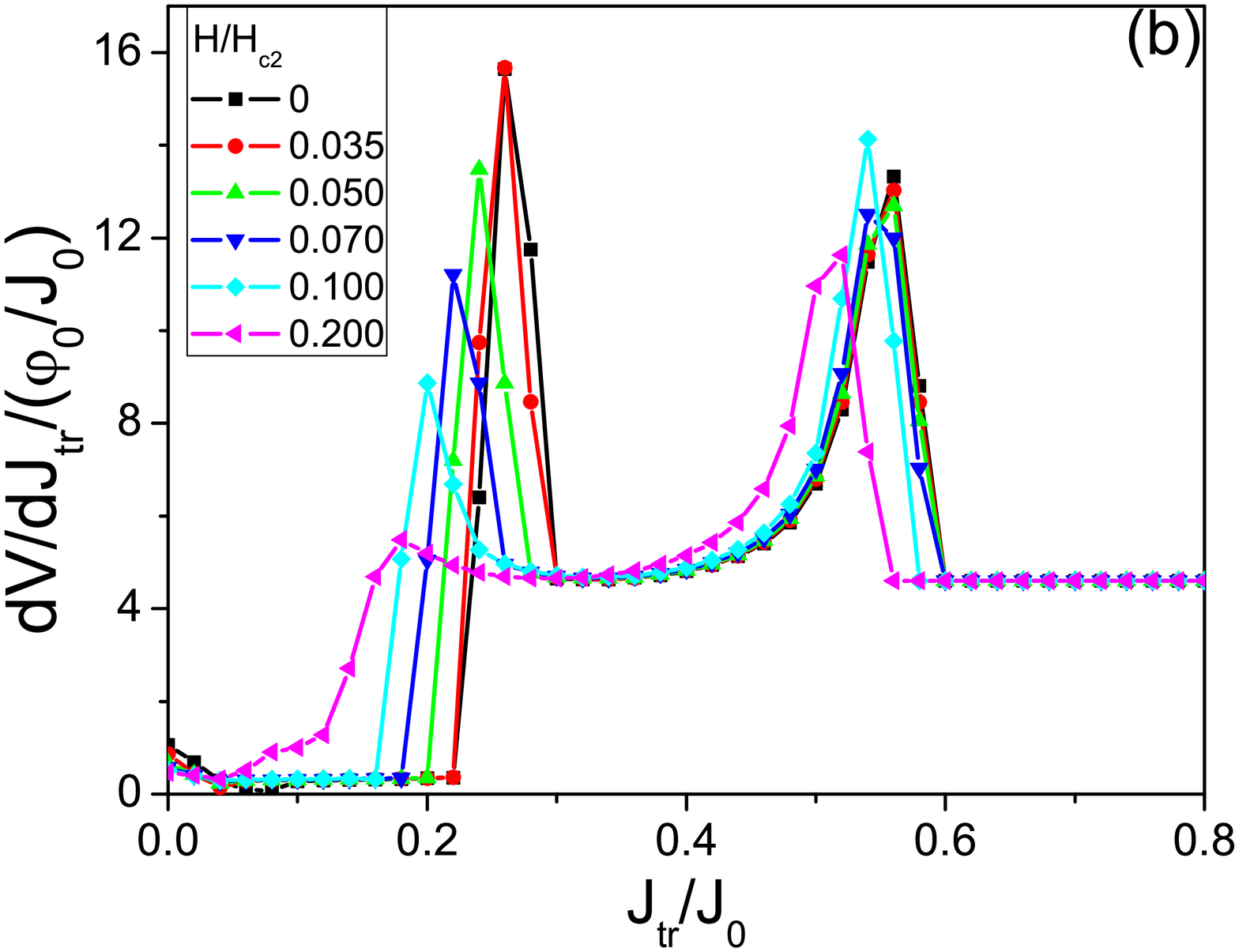}
    \caption{(a) Current-voltage curves to simulated samples. The inset is a zoom of the first jump of the curves, in the cases where it occurs. As the intensity of the applied field varies four different regimes of vortex dynamics exist (see main text for details). (b) Derivative resistance showing two peaks associated with the beginning of the resistive state (lower current) and the transition to the normal state (higher current).}
    \label{fig:curvaIV}
\end{figure}

In panel (a) of Fig.~\ref{fig:curvaIV}, it is presented the normalized characteristic $I-V$ curves for several values of $H$. To avoid ohmic effects due to the electrical contacts, the voltage was calculated between the positions $x=3.8\xi$ and $x=8.2\xi$ at the edge of the sample. The jump in $I-V$ curves, evidenced in the inset, indicates the beginning of the resistive state with the appearance of $kV$ and/or pairs of kinematic vortex and antivortex, depending on the value of $H$. Additionally, panel (b) shows the derivative resistance which, up to intermediate values of the applied field, presents two peaks. The first one, for lower currents, at $J_{c1}$, denotes the beginning of the resistive state and the second peak, $J_{c2}$, the transition to the normal state. As discussed below, for higher values of $H$ a different behavior occurs. Finally, for $H=H_{c2}$ the sample is in the normal state, and one has an ohmic response.   

%Such behavior occurs for all values of $H$. 

According to the vortex dynamics, it is distinguished four field ranges as follow: (i) zero field, (ii) low fields, $H\leq 0.05H_{c2}$, (iii) moderate fields, $0.05H_{c2}<H\leq 0.14H_{c2}$, (iv) and high fields, $H \geq 0.16H_{c2}$. 

\begin{figure}[h!]
    \centering
    \includegraphics[width=4.3cm]{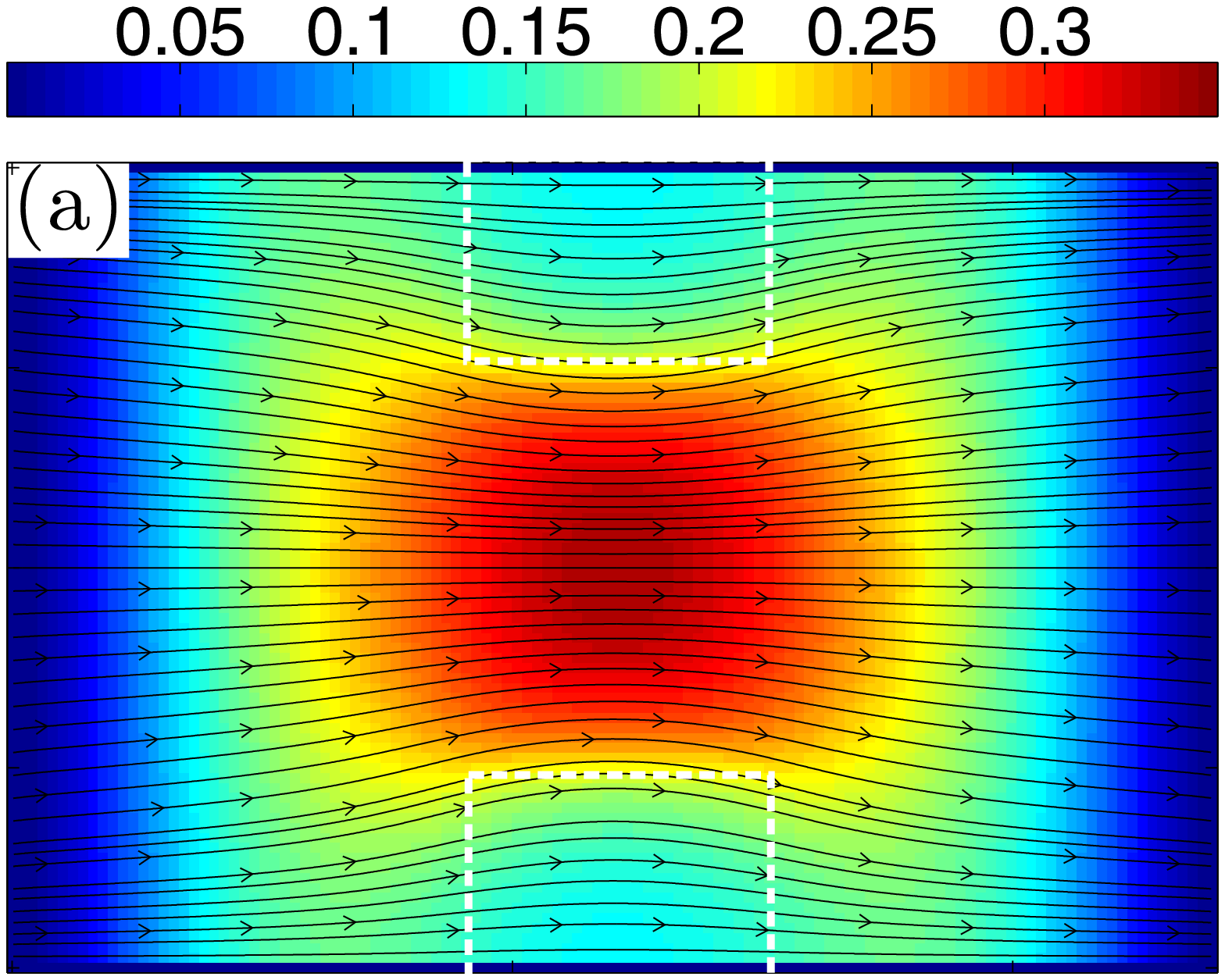}\includegraphics[width=4.3cm]{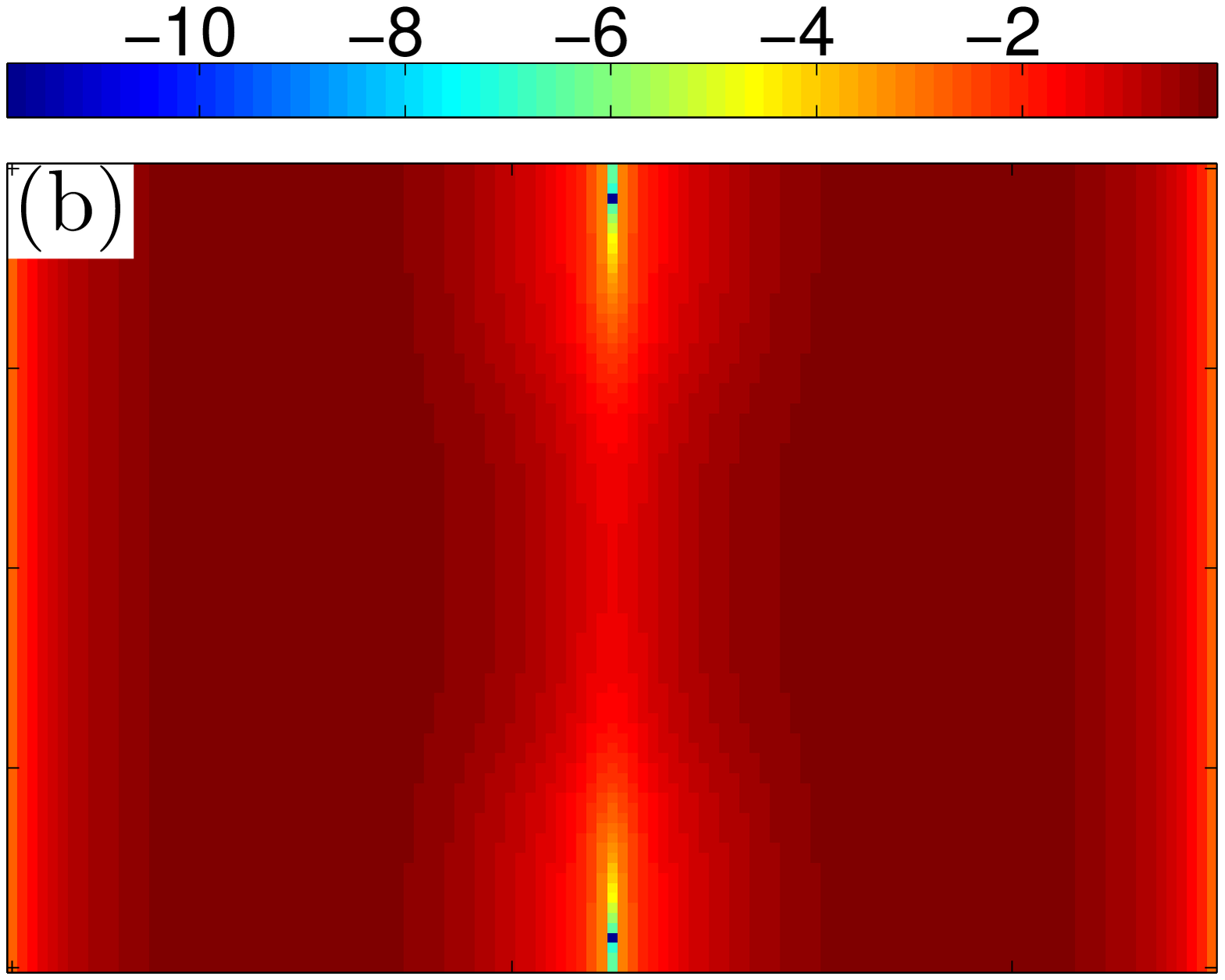} \\
    \includegraphics[width=4.3cm]{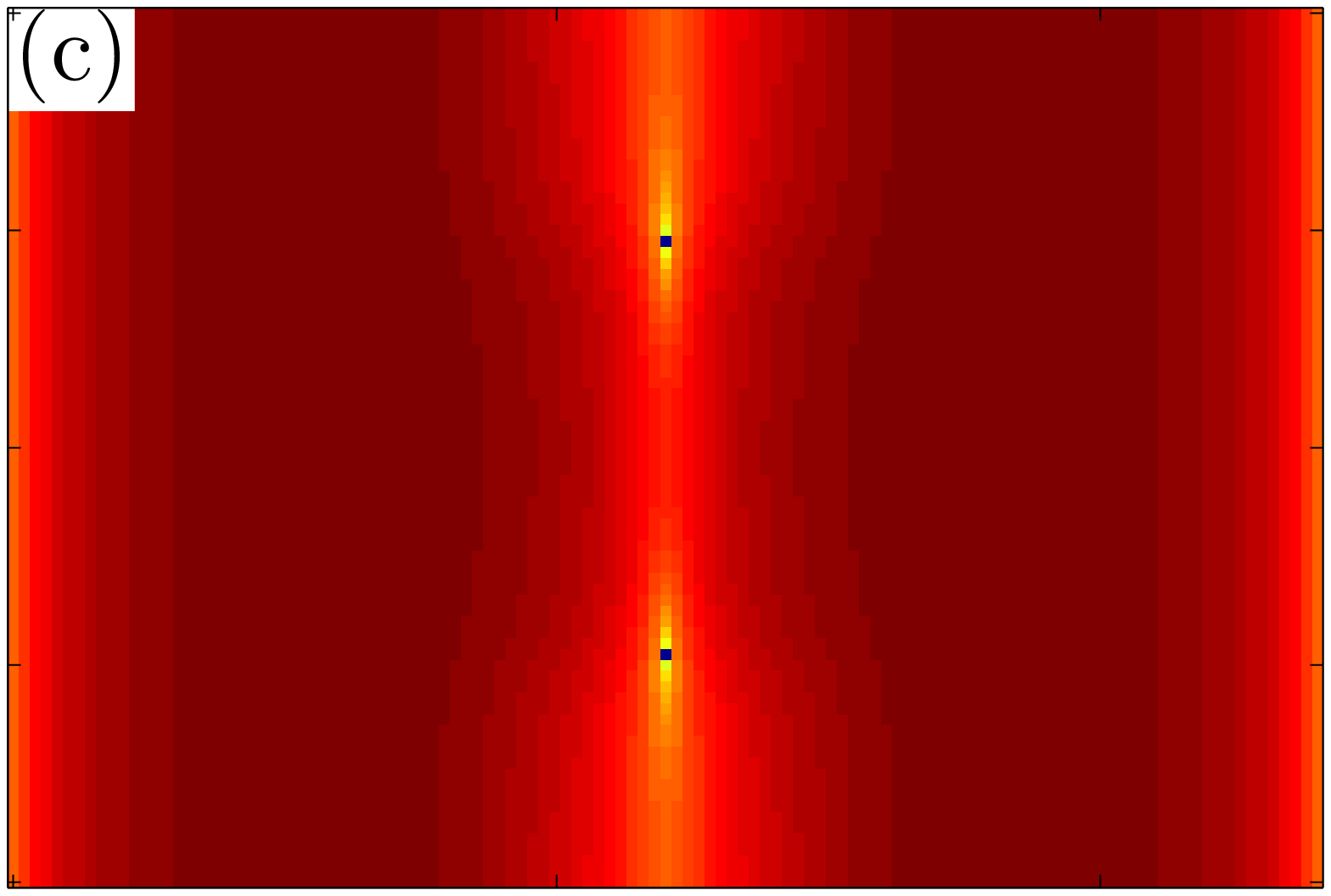}\includegraphics[width=4.3cm]{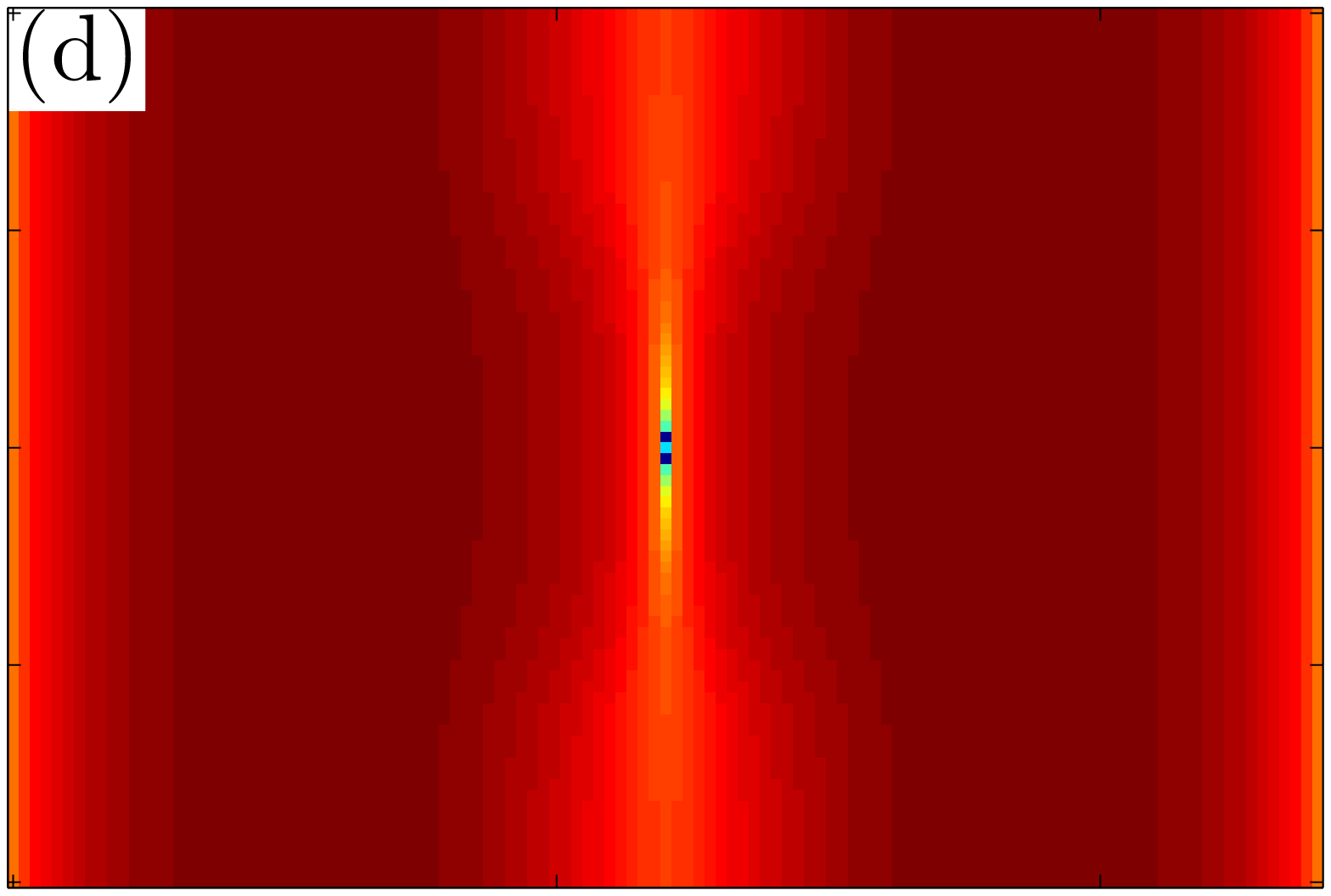}\\
   \vspace{0.35cm}\includegraphics[width=6cm]{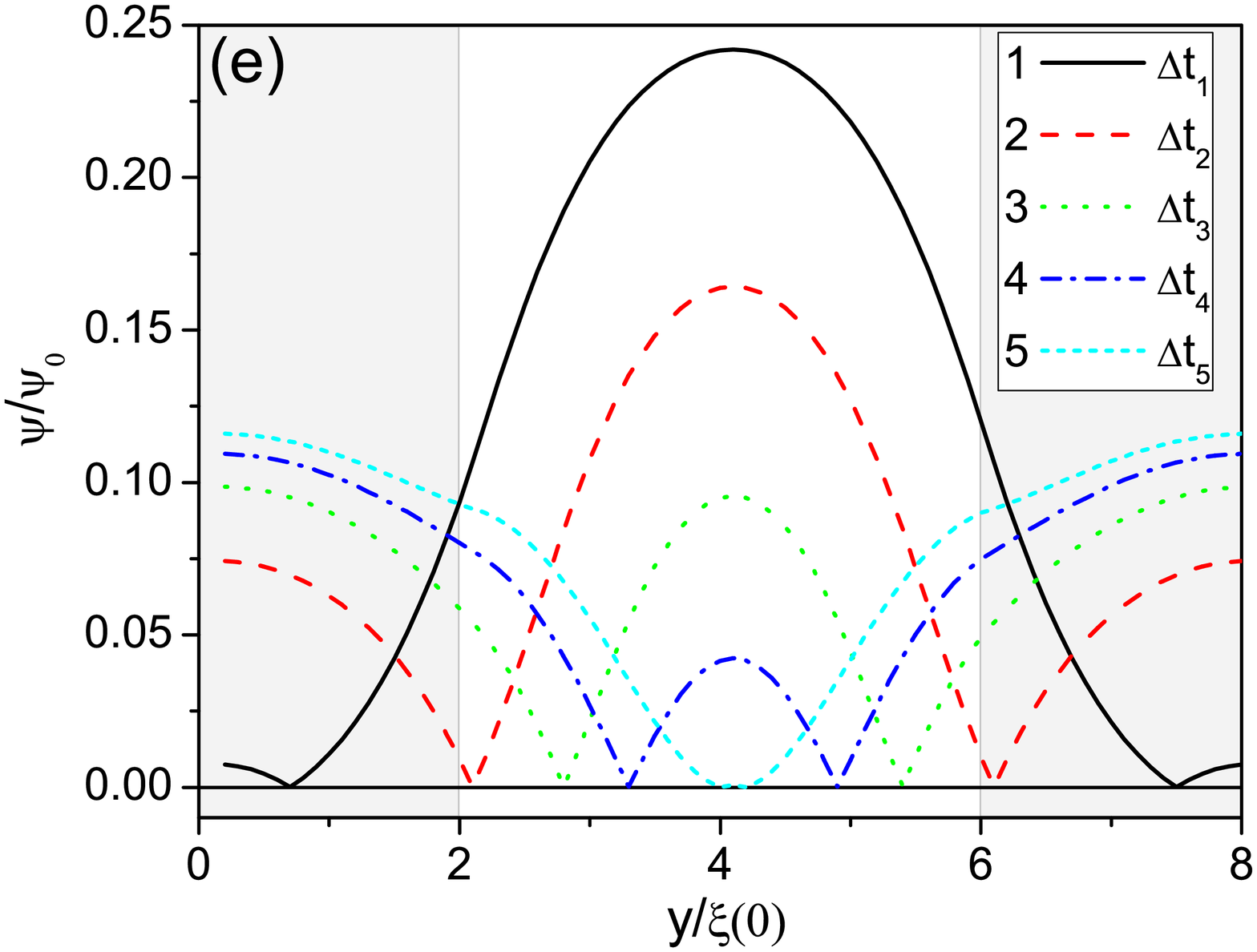}
    \caption{(a) Modulus of $\textbf{J}_s$ superimposed by the streamlines for $H=0$ and $J_{tr}=0.24J_{0}$. The dashed areas represent the borders of the defects. (b) to (d) snapshots of $|\psi|$ in logarithm scale showing the movement of the $V-aV$ pair that culminates in its annihilation at the center of the sample; (e) intensity of $|\psi|$ along the $y$ axis at different times showing the $V-aV$ motion since their penetration until their annihilation.}
    \label{fig:psiHnulo}
\end{figure}

As expected, at $H=0$, $\textbf{J}_s$ is more intense in the constriction, as can be seen in panel (a) of Fig.~\ref{fig:psiHnulo} for $J_{tr}=0.24J_0$. Associated with this, the weaker superconductivity at the defects allows the formation of kinematic $V-aV$ pairs for $J_{tr}\geq J_{c1}$. A pair is formed one each a time and moves to the center of the sample, annihilating themselves. Panels (b) to (d) of Fig.~\ref{fig:psiHnulo} show a sequence of snapshots of the logarithm of $|\psi|$ illustrating the described dynamics. Such dynamics persist until the transition to the normal state. In panel (e) it is shown the intensity of $|\psi|$ along the $y$ axis at different times; the $V$ ($aV$) core center is at $|\psi|=0$, for which the motion from the penetration up to the pair annihilation can be followed. The shaded areas represent the defect regions.
It is worth mentioning that in our case, the system presents only one resistive state, i.e., just one type of $V-aV$ dynamics contrasting to the findings 
described by Berdyiorov \textit{et al}., where two distinct dynamics were verified, for which two resistive states were associated with.\citep{Berdiyorov-Pri} Namely, between $J_{c1}$ and $J_{c2}$, there is a maximum of the resistive curve. At this point, there is an inversion of the collision of the $V-aV$ pair. First, they collide at the center of the sample. Then they are formed at the center and move towards the edges of the sample. It seems that the constriction suppresses this effect. 

\begin{figure}[h!]
    \centering
    \includegraphics[scale=0.9]{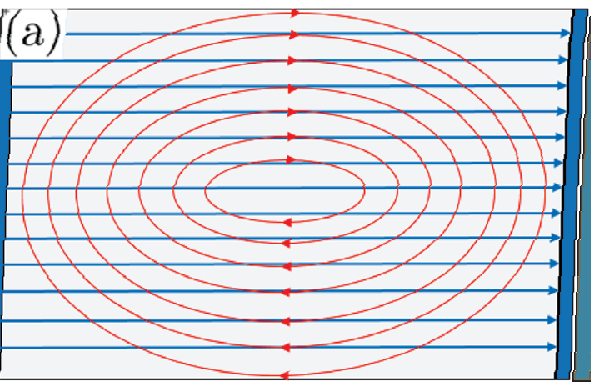}\includegraphics[scale=0.2]{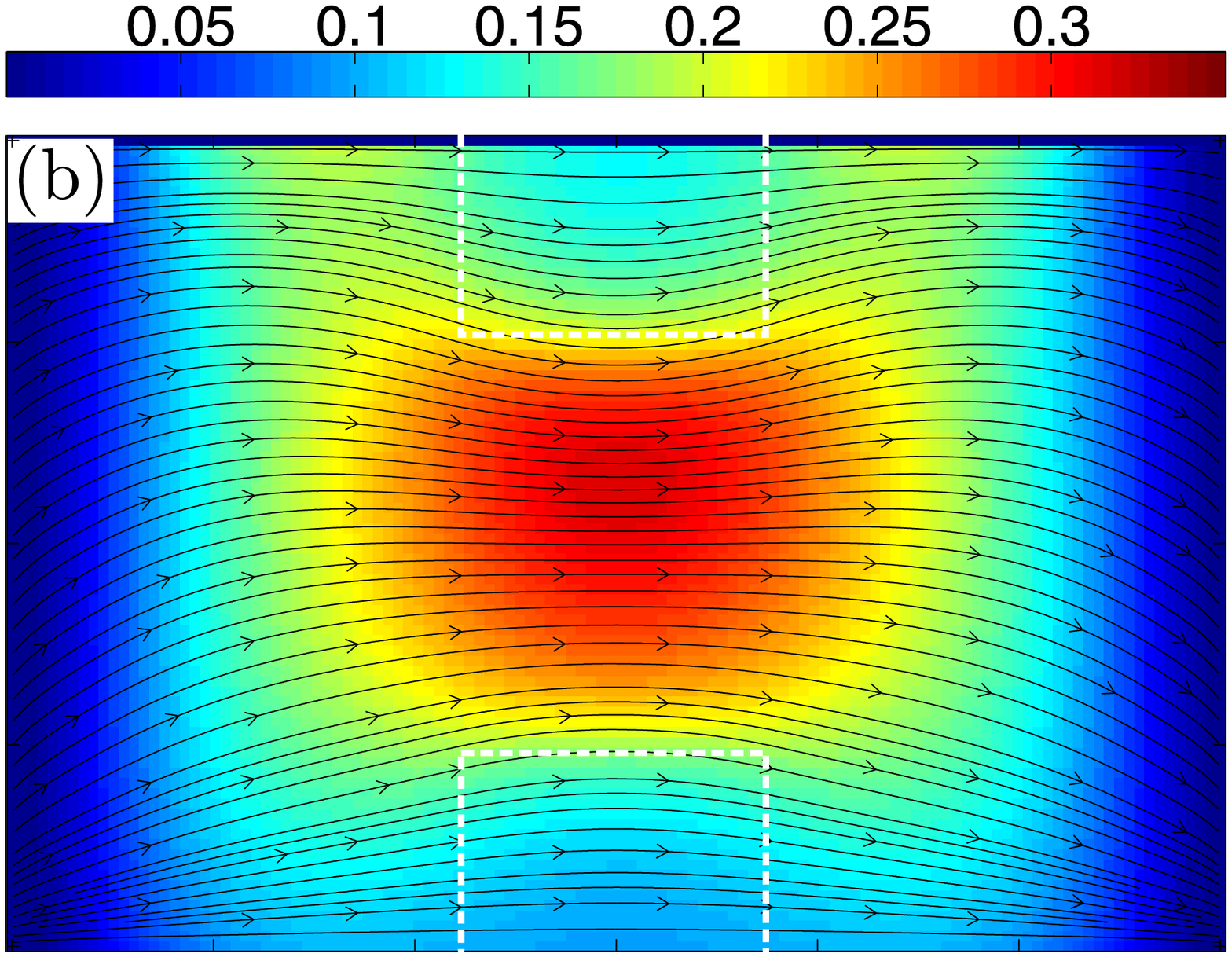}\\
    \includegraphics[scale=0.25]{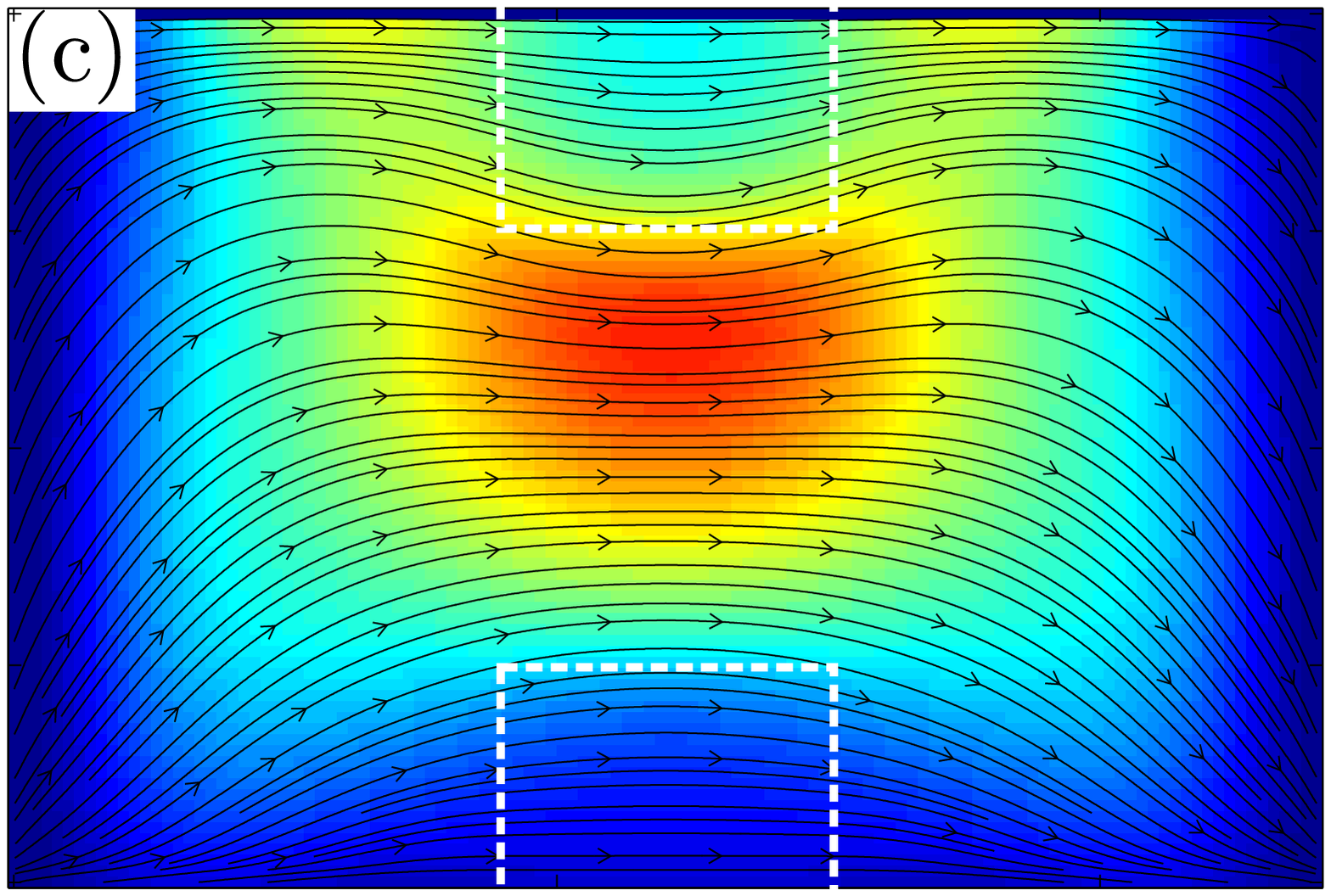}
    \caption{(a) Schematic view of the shielding currents solely due the applied field; (b) and (c) intensity of the modulus of $\textbf{J}_s$ superimposed by the streamlines for $H=0.035H_{c2}$ and $H=0.07H_{c2}$, respectively. The first one for $J_{tr}/J_{0}=0.22$ and the second one for $J_{tr}/J_{0}=0.18$.}
    \label{fig:Hassimetrico}
\end{figure}

Now, the symmetry of the system is broken when an applied magnetic field is considered, causing an asymmetric distribution of $\textbf{J}_s$, as can be observed in Fig.~\ref{fig:Hassimetrico}. Panel (a) of this figure shows a schematic view of the induced superconducting currents only due to $H$, which are circulating. As we apply a transport current, the symmetry of the shielding currents is lost. In panel (b) and (c) we show the map of the modulus of $\textbf{J}_s$ superimposed by the streamlines for $H=0.035H_{c2}$ and $H=0.07H_{c2}$, respectively. Therefore, in the low-field limit (the case of the panel (b) for $J_{tr}=0.22J_{0}$), the $V-aV$ dynamics are the same that those described in Ref. \onlinecite{Berdiyorov-Pri}, i.e., firstly, the $kV$ is formed at the upper edge of the sample and moves straightly towards the lower edge. Just after the appearance of the $kV$, a $kaV$ (kinematic antivortex) is formed, and the annihilation takes place out of the center of the sample. As $H$ increases, the annihilation point approximates to the lower edge.

On the other hand, by increasing $J_{tr}$ at a fixed value of $H$, the formation of the $kaV$ is avoided, and only the $kV$ appears at the upper edge of the sample, leaving it at the lower one. Such behavior depends both on value of $H$ and $J_{tr}$, e.g., at $H=0.035H_{c2}$ the pair is no longer formed for $J_{tr} \ge 0.36J_0$, whereas for the sample at $H=0.05H_{c2}$ the dynamics changes for $J_{tr}\ge 0.28J_0$.

At moderate field regime, the asymmetry in the  distribution of $J_s$ increases, as shown in panel (c) of Fig.~\ref{fig:Hassimetrico} for $J_{tr}=0.18J_{0}$, and a distinct dynamics takes place where no $kaV$ is created. Thus, the resistive state consists of just the appearance of a $kV$ at the upper border of the sample, and it escapes the sample through the lower edge.

\begin{figure}[h!]
    \centering
    \includegraphics[width=8cm]{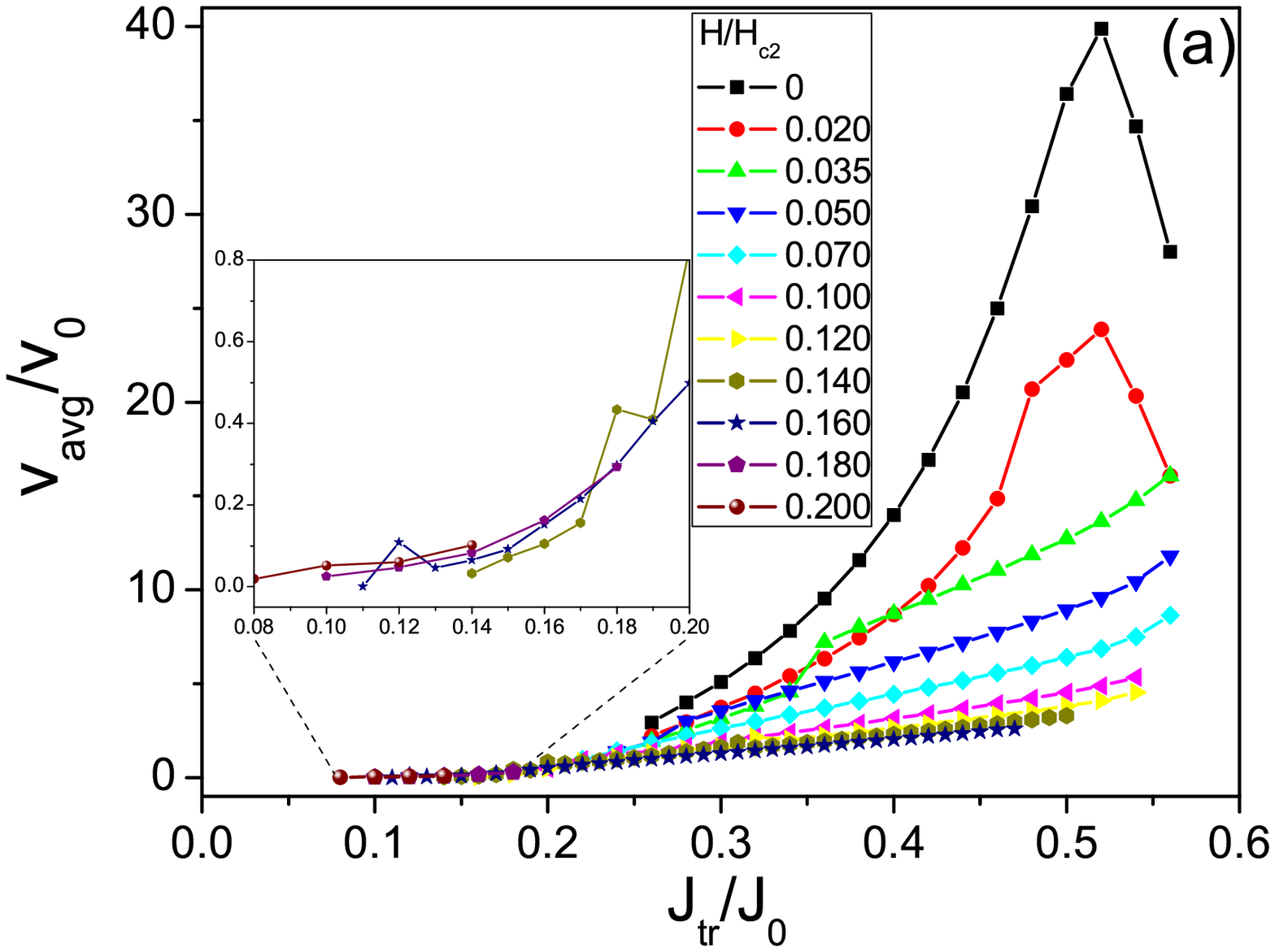}\\
    \includegraphics[width=8cm]{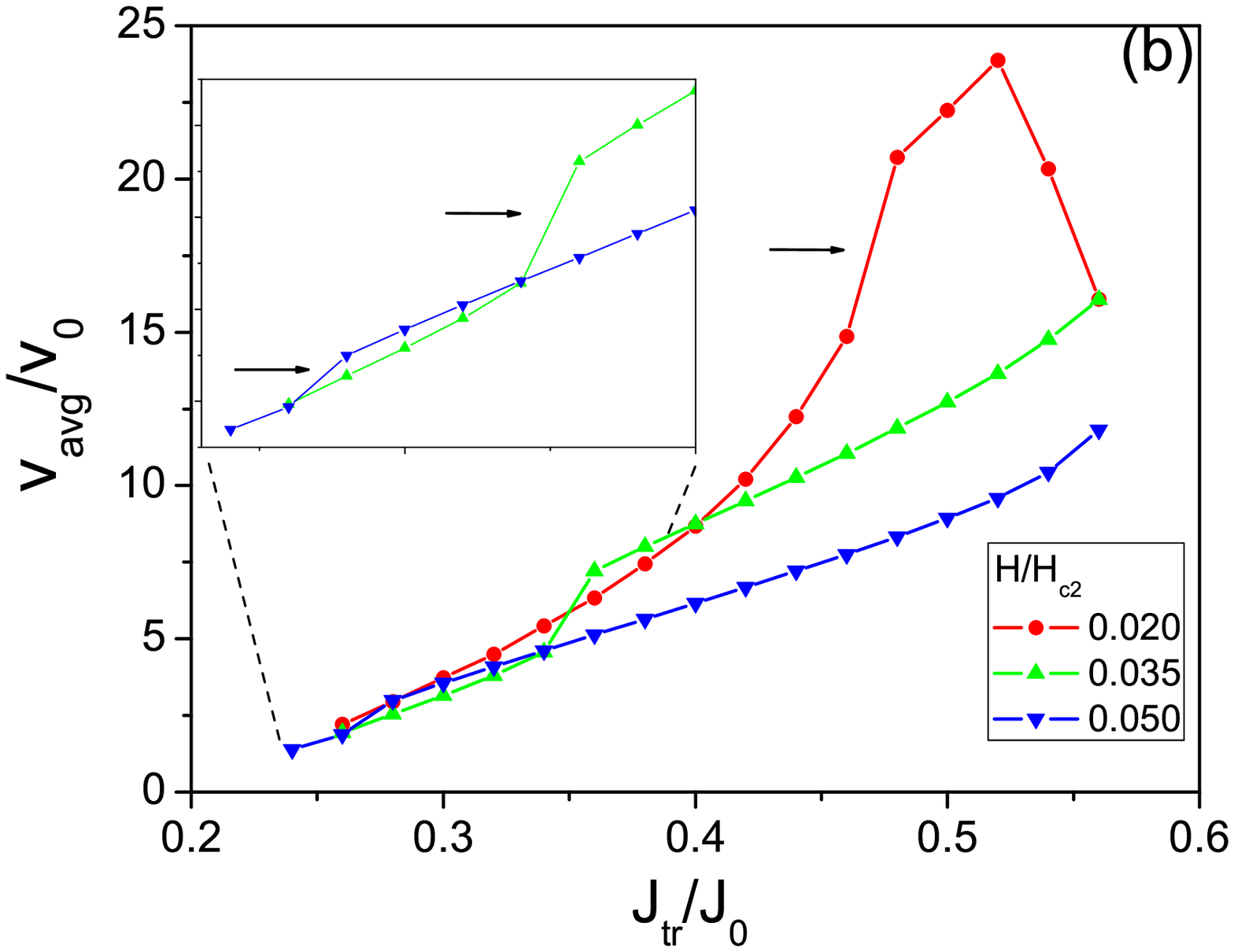}
    \caption{Curve of the average velocity of the vortex as a function of the the applied current. In (a) it may be noted that as the intensity of the field increases, the velocity decreases. In (b) it is evidenced the jump from the curves for the low field regime when there is a change in the dynamics of the vortices.}
    \label{fig:velocidademedia}
\end{figure}

In the high field regime, the velocity of the vortices decreases significantly, near two orders of magnitude, which indicates that the resistive state presented by the $I-V$ curve is caused only by Abrikosov-like vortices in a flux-flow regime. Such $AbV$ velocity is of the same order of magnitude of that one measured in Ref. \cite{Jelic-2016} using a stroboscopic resonance technique. Panel (a) of Fig.~\ref{fig:velocidademedia} exhibits  the average vortex velocity, $v_{avg}$, as a function of $J_{tr}/J_{0}$ for several values of $H$. The inset highlights the high-field curves which show the huge field-dependence of $v_{avg}$, being $10^2$ times smaller then $v_{avg}$ at smaller fields. 

It is worth to mention that it is a tough task to follow the position of a single vortex in the scenario of multiple penetrations of vortices since they are identical particles. Having in mind this difficulty, the high-field curves in Fig.~\ref{fig:velocidademedia} do not reach high values of $J_{tr}$.
On the other hand, the curves related to the low-fields regime present a jump, which is indicated by arrows in panel (b).
Those current jumps evidence the change in the vortex dynamics, i.e., delimit the region where there is the formation of only a $kV$, which moves across the sample until it leaves through the opposite side.

\subsection{\label{sec:level13b} Amplitude, Frequency and $J_{tr}(H)$ Phase Diagram}

Now, we analyze the frequency and amplitude of the time-voltage characteristic curves, since they might be experimentally observed.\cite{PhysRevLett.91.267001} As we vary the magnetic field applied to the superconducting sample, we noticed that the frequency and amplitude as functions of $J_{tr}$ present different behaviors for each field regime described previously, i.e., low, moderate and high fields. The same occurs with the time-voltage characteristic curves.

\begin{figure}[h!]
    \centering
    \includegraphics[width=8.5cm]{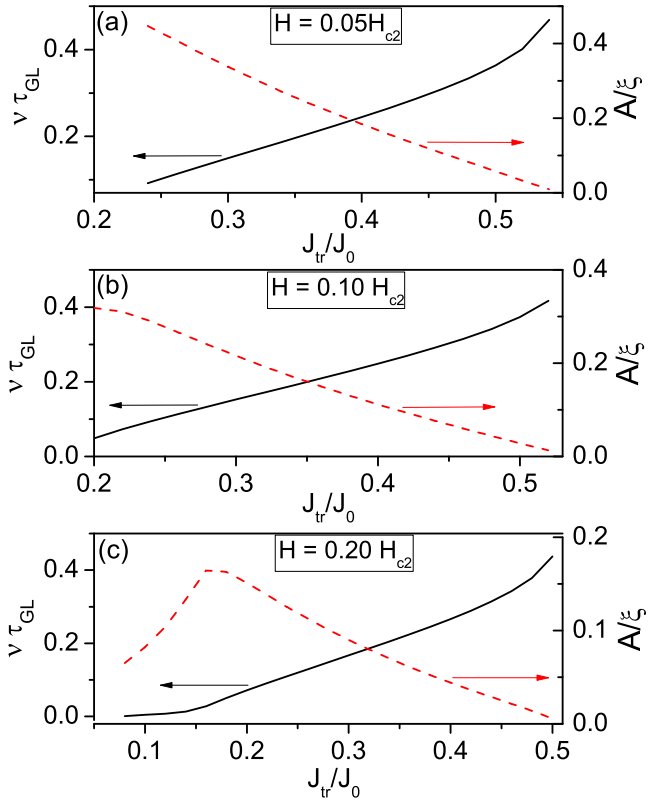}
    \caption{The frequency and the amplitude as functions of the applied current for three different values of the applied field: (a) $H=0.05H_{c2}$, (b) $H=0.1H_{c2}$, and (c) $H=0.2H_{c2}$. It can be noticed that, as the intensity of the applied field increases, the curves present different behaviors, especially regarding the amplitude.}
    \label{fig:FreqAmpli}
\end{figure}

Panels (a), (b)  and (c) of Fig.~\ref{fig:FreqAmpli} show the frequency and amplitude for the applied fields $H=0.05H_{c2}$, $H=0.1H_{c2}$ and $H=0.2H_{c2}$, respectively. Observing the amplitude curves (dashed lines), we notice that, initially, $A$ decreases monotonically from low to moderate field regimes. As $H$ increases, the curve tends to develop a maximum (see panel (c)). On the other hand, the frequency presents a smooth growth in all cases.

\begin{figure}[h!]
    \centering
    \includegraphics[width=8.5cm]{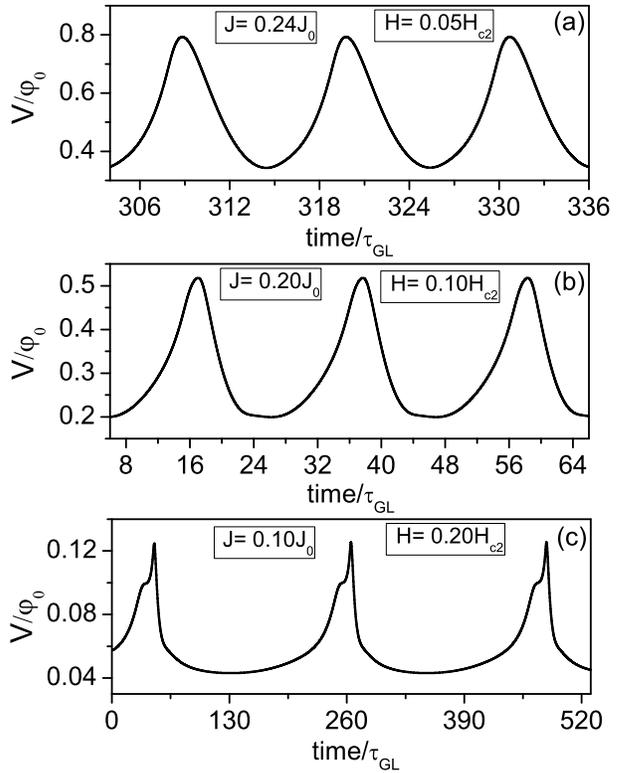}
    \caption{Time-voltage characteristic curve three different values of applied field: (a) $H=0.05H_{c2}$, (b) $H=0.1H_{c2}$, and (c) $H=0.2H_{c2}$. We see that, by increasing $H$, the oscillations present a deformation and the period increases, due to the lower velocity of the vortex across the sample.}
    \label{fig:tv}
\end{figure}

By carefully inspecting the time-dependence of the voltage (see Fig.~\ref{fig:tv} which corresponds to the same values of $H$ as in Fig.~\ref{fig:FreqAmpli}), we can see a considerable increase of the period of the events as a consequence of the decreasing of the vortex velocity with increasing $H$. Additionally, the oscillations gradually became a pulse-like signal by increasing $H$ (see panel (c)).

\begin{figure}[h!]
    \centering
    \includegraphics[width=8.5cm]{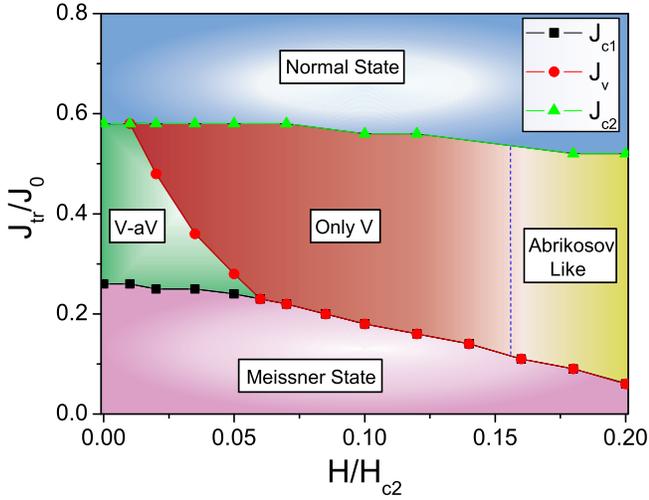}
    \caption{The $J_{tr}(H)$ phase diagram. Just above $J_{c1}(H)$ the resistive state sets in; the region comprehended in between $J_{c1}$ and $J_v(H)$ only the $V-aV$ pairs are formed, and above $J_v(H)$ up to $J_{c2}(H)$ only vortices exist. Finally, above $J_{c2}(H)$ the system goes to the normal state. In the supplementary material, it can be seen videos corresponding to the dynamics presented here.}
    \label{fig:jxh}
\end{figure}

Such rich dynamics presented by the $kV$ were compiled in a $J_{tr}(H)$ phase diagram shown in Fig.~\ref{fig:jxh}. The $J_{c1}(H)$ curve indicates the current for which a resistive state begins. The domain comprehended between the $J_{c1}(H)$ and $J_v(H)$ curves, it is where it occurs the annihilation between a $kV$ and a $kaV$. Outside such a region, there is only the formation of $kV$'s, which experiences a surface-barrier effect as it approximates the lower border of the sample (see next Section~\ref{sec:level3c}). The dashed line indicates the beginning of a flux-flow regime with Abrikosov-like vortices. The $J_{c2}(H)$ curve delimits the transition to the normal state.

\begin{figure}[h!]
    \centering
    \includegraphics[width=8.5cm]{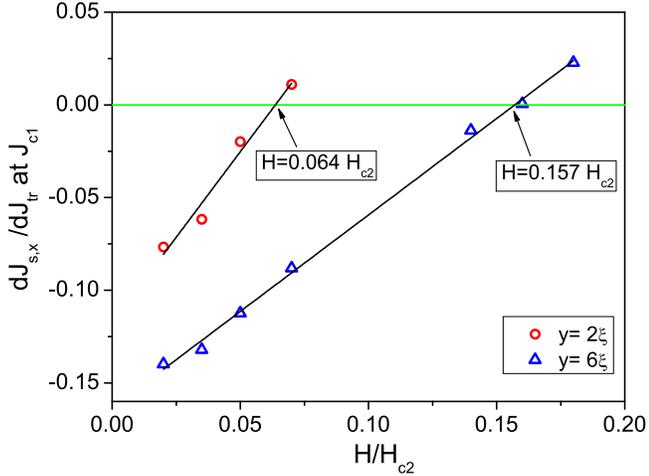}
    \caption{Derivative of $J_{s}$ horizontal component with respect to $J_{tr}$ at $J_{c1}$ is plotted as function of $H$ for $y=2\xi$ and $y=6\xi$. Highlighted $H$ indicate the end of $kaV$ formation at $y=2\xi$ and $kV$ at $y=6\xi$.} 
    \label{fig:dJs/Jtr}
\end{figure}

In order to give a more quantitative consistency to the $J_{tr}(H)$ phase diagram, we analyzed the behavior of the current density, both at the upper and lower edge of the constriction. Our argument is as follows. The lines connecting points in Fig.~\ref{fig:jxh} are a guide to the eyes. However, a quantitative estimation of two qualitatively important points in that diagram is offered in Fig.~\ref{fig:dJs/Jtr}, where the derivative of $J_{s}$ horizontal component (denoted by $J_{s,x)}$) with respect to $J_{tr}$ at $J_{c1}$ is plotted as function of $H$ at the lower ($y=2 \xi$) and upper ($y=6 \xi$) borders of the constriction for $x=6 \xi$ (the center of the sample). Those points in the sample are in the intersection between the kinematic $V-aV$ path and the defect borders. By comparing Figures \ref{fig:jxh} and ~\ref{fig:dJs/Jtr} one sees that the nucleation of $kaV$ and $kV$ cease at values of $H$ for which the mentioned derivative changes its sign at $y=2\xi$ and $y=6\xi$, respectively. It is important to recall that the $kaV$ always nucleates at the lower border of the sample, and the $kV$ always at the upper border. This behavior signals how strongly the interplay between $J_{tr}$ and the shielding currents (the latter being controlled by $H$) affects the kinematic $V-aV$ dynamics.

\subsection{\label{sec:level3c} Effects of Surface Barrier on Kinematic Vortex}

\begin{figure}[h!]
    \centering
    \hspace{-0.5cm}\includegraphics[scale=0.35]{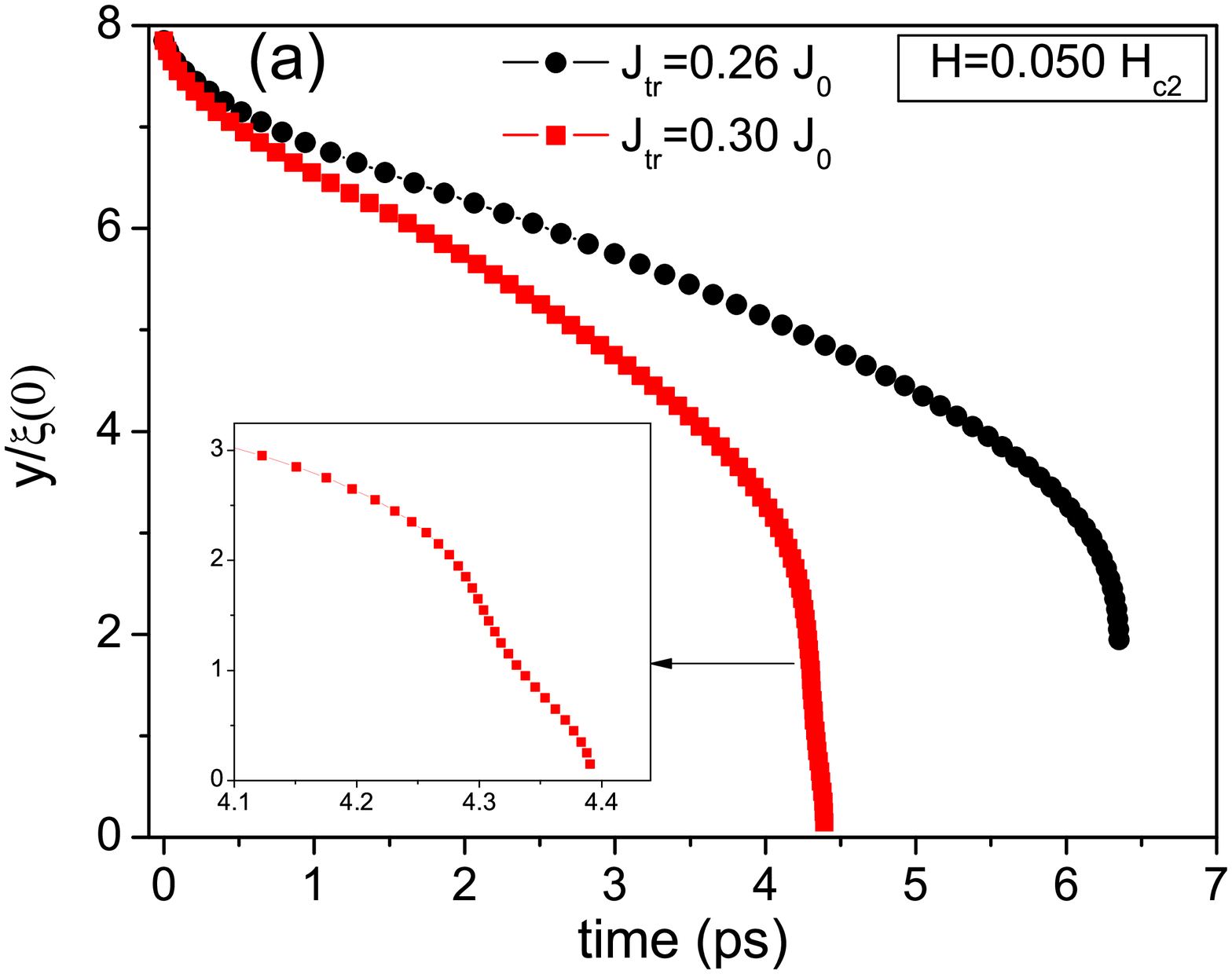}\\
    \hspace{-0.5cm}\includegraphics[scale=0.35]{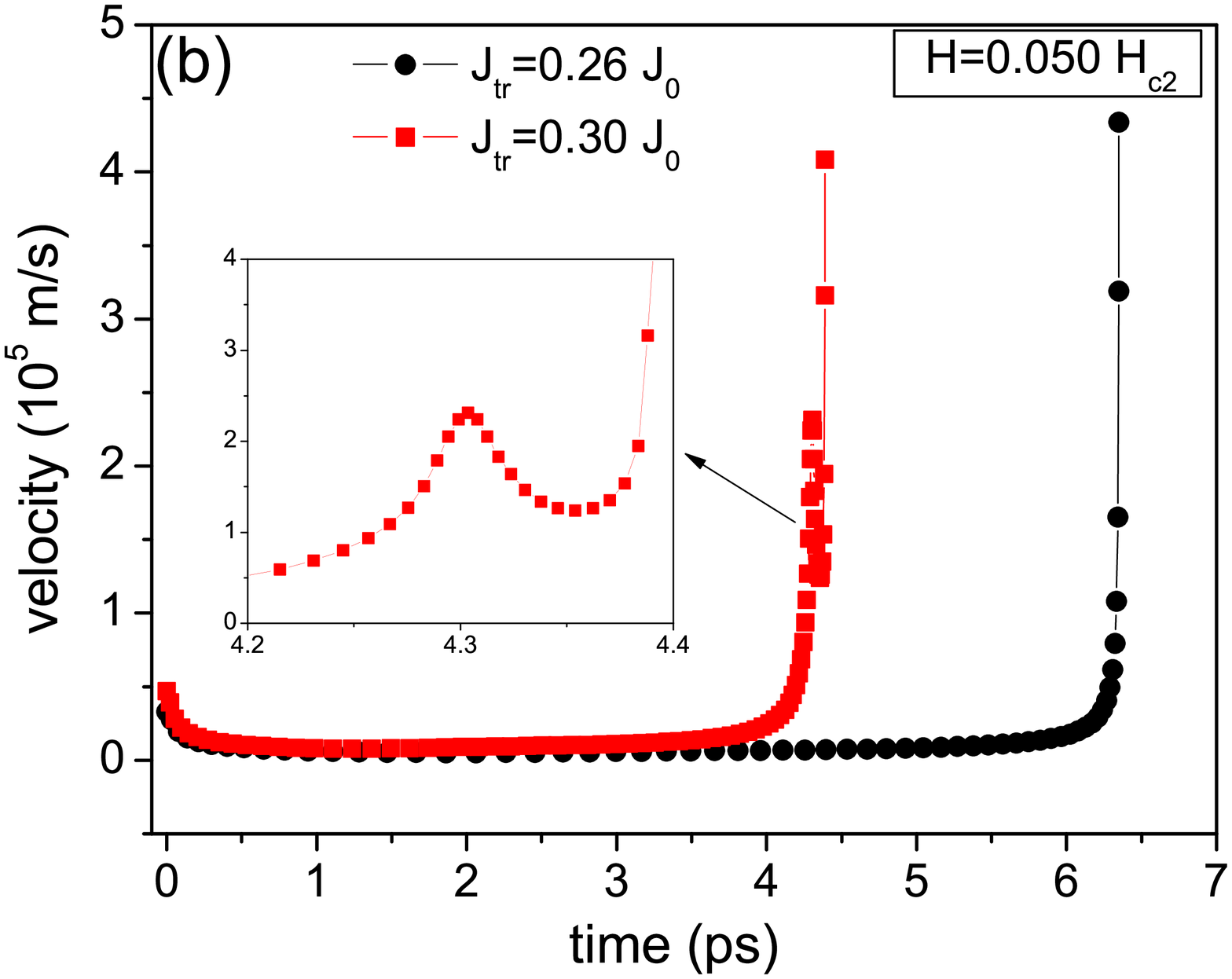}\\
    \includegraphics[scale=0.35]{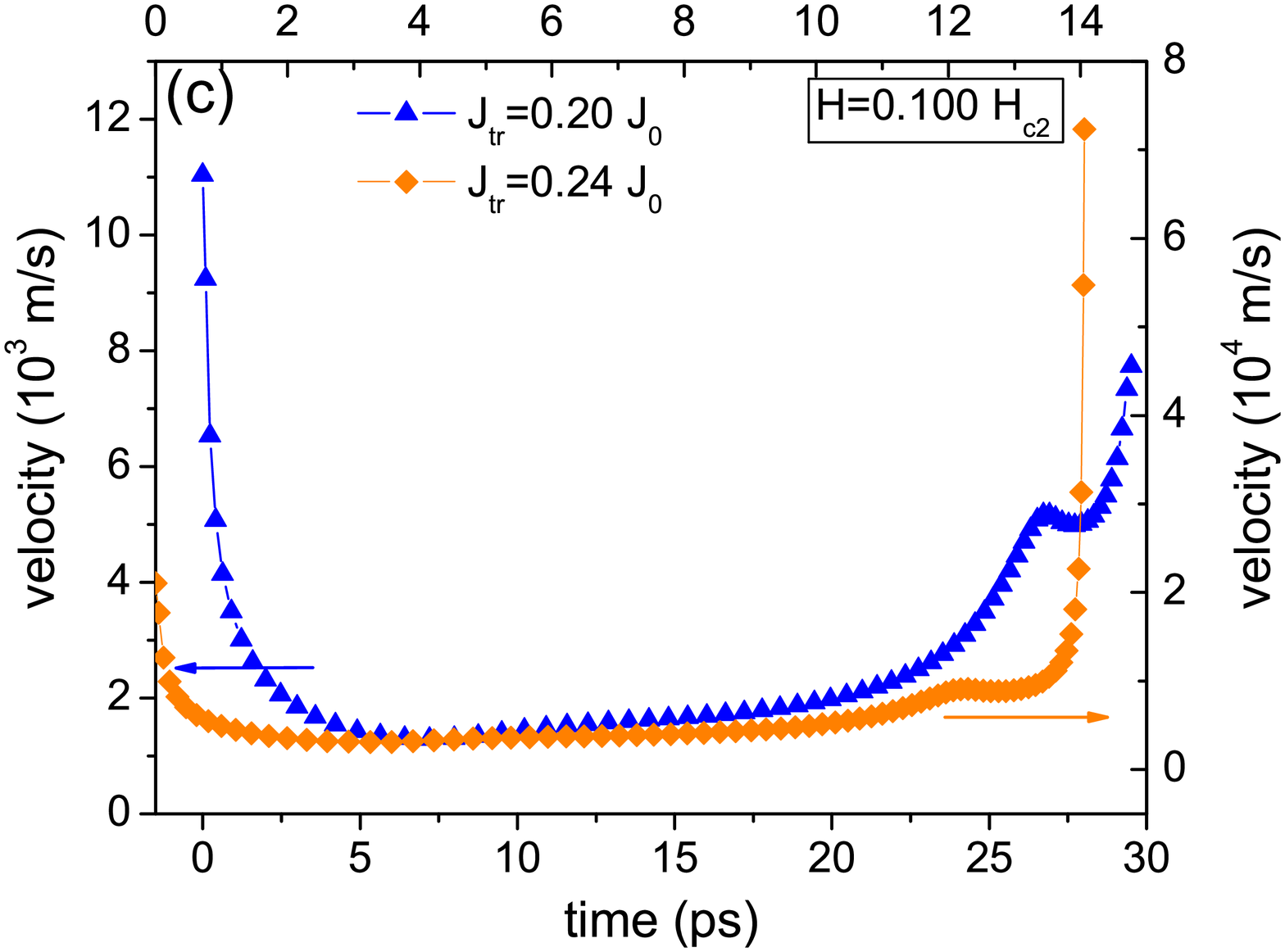}
    \caption{(a) Vertical vortex position as a function of the time for $H=0.05H_{c2}$ and two different regimes of vortex dynamics. The inset shows a time-changing of the $V$ motion beginning when it is passing by the edge of the defect ; velocity curve as a function of time for (b) $H=0.05H_{c2}$, in the black curve without the effect of surface barrier and in the red curve surface barrier effect decreasing the speed; (c) $H=0.1H_{c2}$ and two values of $J_{tr}$, showing that the  surface barrier effectiveness decreases with increase of $J_{tr}$.}
    \label{fig:velxtempo}
\end{figure}

As mentioned previously, in this Section, we will analyze the effect and origin of the surface barrier. Particularly, the system at $H=0.05H_{c2}$ presents two distinct dynamics. One of them is characterized by the annihilation of a $kV$ and a $kaV$. In such a process, the $kV$ accelerates at the beginning of its motion, and during the annihilation (see the circles in Fig.~\ref{fig:velxtempo} (a) and (b)). As $J_{tr}$ increases, the $kaV$ is no longer formed, and then, the $kV$ experiences a decreasing of its velocity as it approximates the lower border of the sample, (see the squares in Fig.~\ref{fig:velxtempo}(a) and (b)). At those curves, we used real units by considering the parameters of a Nb sample,\citep{Gubin-2005} i.e., $T_c=9.2K$ and $\xi(0)=10$ nm to illustrate the time range of the $kV$ motion. As can be seen in the insets of panels (a) and (b) of that same figure, the surface barrier is effective just during a very short period of time (of the order of picoseconds), where the velocity of the $kV$ begins to reduce in the vicinity of the frontier between the superconducting matrix and the defect. After that, the interaction with its vortex-image accelerates the $kV$ once again, leaving the sample with high velocity. In this way, as in nanometric superconductors, the surface takes a major role in their dynamics. Then, the knowledge about such a noteworthy mechanism could be useful to control dissipating processes due to $kV$ and phase-slips.

The effectiveness of the surface barrier effect decreases due to the increase of the Lorentz force for higher $J_{tr}$. As a consequence, the $kV$ moves faster (increasing its kinetic energy) to leave the sample. This behavior is shown in Fig.~\ref{fig:velxtempo}(b) for the system at $H=0.1H_{c2}$. The curve with triangles shows the velocity as a function of time at $J_{tr}=0.20 J_0$, and the one with diamonds at $J_{tr}=0.24J_0$. We see in Fig.~\ref{fig:velxtempo}(b) that with increasing the current, the velocity can be one order of magnitude as large as much, and consequently, the surface barrier effectiveness is reduced.

\section{\label{sec:level4}Conclusions}

In this work, we simulated a superconducting stripe with a central constriction by varying the $dc$ applied transport current at fixed applied magnetic fields. We show that the kinematic vortex-antivortex dynamics are quite rich, which is evidenced in the $J_{tr}(H)$ diagram of Fig.~\ref{fig:jxh} exhibiting a variety of superconducting phases. It is remarkable that despite occurring in a very short time range (of the order of picoseconds), the kinematic vortex experiences the effects of a surface barrier when in the absence of kinematic antivortex formation. Since resistive states are not desirable for several applications of mesoscopic superconductors, we claim that the detailed knowledge about vortex dynamics is a crucial element to avoid them. Therefore, our results could contribute to future applications and the design of devices.

\section{\label{sec:level5}Acknowledgments}
We acknowledge the Brazilian agencies S\~ao Paulo Research Foundation (FAPESP, Grant 2016/12390-6), Coordena\cao de Aperfei\c coamento de Pessoal de N\ii vel Superior - Brasil (CAPES) - Finance Code 001, and National Council of Scientific and Technological Development (CNPq, grant 302564/2018-7).

\bibliographystyle{apsrev4-1}
\bibliography{refs.bib}

\end{document}